\documentclass[12pt,twoside]{article}
\usepackage{amsmath,amsfonts,amssymb}
\usepackage{latexsym}
\usepackage{dcolumn}
\usepackage{graphicx,epsfig}
\usepackage{amsthm}
\usepackage{hyperref}

\begin{document}

%\begin{flushright}
%{\tt IIT Gandhinagar/GR-QC\\ \today}
%\end{flushright}
%\vspace{1.5cm}

\begin{center}
{\Large \bf The Effects of Minimal Length in Entropic Force Approach}
\vglue 0.5cm
Barun Majumder\footnote{barunbasanta@iitgn.ac.in}
\vglue 0.6cm
{\small Indian Institute of Technology Gandhinagar \\ Ahmedabad, Gujarat 382424}
\end{center}
\vspace{.1cm}

\begin{abstract} 
With Verlinde's recent proposal which says that gravity can be identified with an entropic force and considering the effects of generalized uncertainty principle in the black hole entropy-area relation we derive the modified equations for Newton's law of gravitation, modified Newtonian dynamics and Einstein's general relativity. The corrections to the Newtonian potential is compared with the corrections that come from Randall-Sundrum II model and an effective field theoretical model of quantum general relativity. The effect of the generalized uncertainty principle introduces a $\sqrt{\text{Area}}$ type correction term in the entropy area relation whose consequences in different scenarios are discussed.
\vspace{5mm}\newline Keywords: entropic gravity, generalized uncertainty principle
\end{abstract}
\vspace{1cm}

%START
\section{Introduction}

One of the greatest achievements in theoretical physics is the realization that black holes are well defined thermodynamic objects with entropy and temperature \cite{ha1,ha2,ha3}. Hawking \cite{ha3} has derived that a Schwarzschild black hole emits a thermal radiation whose temperature depends on the mass $M$ of the black hole and is given by $T=\frac{1}{8\pi M}$. Also Bekenstein has shown that a black hole has a well defined entropy and is proportional to the area of the black hole horizon given by the entropy area relation
\begin{equation} 
\label{areabek}
S_{BH} = \frac{A}{4 l_p^2} ~.
\end{equation}
Here $A$ is the cross sectional area of the black hole horizon and $l_p$ is the Planck length. Recently there has been much interest devoted to the leading order quantum corrections of the black hole entropy area relation. Entropy accounts for the number of microstates of the system as it has a definite statistical meaning in thermodynamics. A. D. Sakharov is the originator of the idea of emergent gravity \cite{sakh}. Jacobson \cite{s2} was the first to view Einstein's equation as an equation of state. Together with the second law of thermodynamics and the fact that entropy is proportional to the horizon area he derived the Einstein's equations. Later several studies were carried out to understand the deeper underlying connection between horizon thermodynamics and Einstein's equation. Padmanabhan showed that for a wider class of theories the gravitational field equations on the horizon can be reduced to the first law of thermodynamics arguing the Einstein's equation to be a thermodynamic entity \cite{s3}. This novel idea was also later introduced in modified theories of gravity \cite{s4}. For a brief review on the demonstration of the idea in other scenarios we refer to \cite{s13,s5}. The development in the lines discussed here refers to the point that thermodynamic properties can be associated to the horizon and gravity can be thought of as an entity whose origin is statistical in nature.\par 
Recently Verlinde \cite{s14} introduced a very interesting proposal to understand the thermodynamic origin of gravity. According to him the changes in information which is associated among material bodies is the prime cause of gravity which is an entropic force. This even demands an explanation of the Newton's law of inertia and the equivalence principle may suggest that the origin of the law of inertia is entropic in nature. In his approach Newton's second law of motion can be recovered if one considers the idea of entropic force which is an effective macroscopic force which originates due to the statistical tendency of the increase of entropy. Also we have to consider the Unruh temperature which is the temperature experienced by an observer in an accelerated frame ($T=\frac{\hbar a}{2\pi k_B c}$). Another observation is the recovery of the Newton's law of gravitation and its relativistic generalization to the Einstein's equation. For that the approach considers the idea of entropic force along with the equipartition of energy and the holographic principle. Though a thermodynamic interpretation of gravity can be given with the equipartition argument even in non relativistic limit was established earlier \cite{s16}. Many authors recently focused on the understanding of the entropic force and the references \cite{s18} outline the literature.\par
In Verlinde's formalism he defined the Newton's constant $G$ through the relation
\begin{equation}
N = \frac{Ac^3}{G\hbar}
\end{equation}
where $N$ is the total number of bits and this relation follows from the holographic principle. Although he showed that this $G$ can actually be related to the Newton's constant. Following the holographic principle it is a natural assumption that the number of bits is proportional to the area $A$. In a theory of emergent space area is defined in this form. Now as $l_p^2 = \frac{G\hbar}{c^3}$ where $l_p$ is the Planck length we get $N=\frac{A}{l_p^2}$. As the Bekenstein-Hawking entropy area relation is $S_{BH}=\frac{A}{4l_p^2}$ we have $N=4S$.\par
Different theories of quantum gravity (e.g., \cite{my25,my26,my29,my30,my31}) have predicted the following form for the entropy of a black hole:
\begin{equation}
S = \frac{A}{4l_p^2} + c_0 \ln \left( \frac{A}{4l_p^2} \right) + \text{const}.
\end{equation}
$c_0$ is a model dependent parameter and $l_p$ is the Planck length. The speculation that the Heisenberg's uncertainty principle could be affected by the presence of gravity was done by Mead \cite{new19}. In the strong gravity regime, conventional Heisenberg uncertainty relation is no longer satisfactory (though approximately). Later modified commutation relations between position and momenta commonly known as the Generalized Uncertainty Principle (GUP) were proposed by string Theory, Doubly Special Relativity (DSR) and black hole physics with the prediction of a minimum measurable length \cite{new20,new21,new22,new23,new24,new25,new26,new27,new28,new29}. Similar kind of modification can also be found in the context of Polymer Quantization in terms of polymer mass scale \cite{new30}. Importance of the GUP can also be realized on the basis of simple gedanken experiments without any reference of a particular fundamental theory \cite{new27,new28}. So the GUP can be thought of as a model independent proposal, ideally suitable for the investigation of black hole entropy. The authors in \cite{new31} proposed a GUP which is consistent with DSR, string theory and black hole physics. This GUP is approximately covariant under DSR transformations but not Lorentz covariant \cite{new29}. With the GUP as proposed in \cite{new31} we can arrive at the corrected entropy area relation for a black hole which can be written as \cite{came,new37,new}
\begin{align}
\label{ent}
S  ~\simeq & ~\frac{A}{4 l_p^2} + \alpha~ \sqrt{\frac{A}{4 l_p^2}} + \beta~ \ln \left(\frac{A}{4 l_p^2}\right) \nonumber \\
& + \sum_{m=\frac{1}{2},\frac{3}{2},\ldots}^\infty \gamma_m  \left(\frac{A}{4 l_p^2}\right)^{-m} + 
\sum_{n=1,2,\ldots}^\infty \delta_n  \left(\frac{A}{4 l_p^2}\right)^{-n} + \mathit{const.} ~~.
\end{align}
This is by far the most general form of quantum corrected entropy area relation. In \cite{came} black hole thermodynamics was first studied with modified dispersion relations and generalized uncertainty principle. Recently, many authors have suggested \cite{add45,add46} that the GUP implications can be measured directly in tabletop experiments which will definitely confirm the theoretical predictions of some models. If not everything then also we can get some experimental bound on the deformation parameters $\alpha$ and $\beta$.
\par
So in this paper we will study the effect of this corrected entropy area relation in the theory of modified Newtonian dynamics (MOND). We will also follow Verlinde's viewpoint to construct the modified Newton's law and Einstein's equation with the entropy corrected relation of (\ref{ent}). $\alpha$ and $\beta$ are model dependent parameters in eqn.(\ref{ent}) and there are some predicted signs and values for $\alpha$ and $\beta$. But here we will consider a general treatment without concentrating on the values for the parameters. A similar approach was carried out by authors in \cite{ran1,s} where they considered logarithmic correction to the entropy area relation \cite{my25,my29,my31} and the power law corrections \cite{s3940}. In \cite{nozari} the effect of GUP on the Newton’s law is studied in a different approach.

\section{Entropic Corrections due to GUP and the Modified Newtonian Dynamics (MOND)}

In 1983 Milgrom \cite{s41} gave a proposal to modify the Newtonian dynamics (commonly known as MOND) which can act as an alternative to non-baryonic dark matter. After realizing the mass discrepancies in the galaxy rotation curves he proposed that for acceleration smaller than $1.2 \times 10^{-10}~m/s^2$ Newtonian dynamics needs a modification. Asymptotically the acceleration due to gravity is $a = \sqrt{a_N~a_0}$ where $a_N$ is the Newtonian acceleration and $a_0=1.2 \times 10^{-10}~m/s^2$. MOND cannot be tested within the solar system as the strong gravitational field of the Sun dictates the dynamics. Usually the MOND acceleration due to gravity $a$ is written as 
\begin{equation}
a_N = a \mu \left(\frac{a}{a_0}\right) ~~.
\end{equation}
In the asymptotic limit the interpolation function $\mu (\frac{a}{a_0})$ admits $\mu =1$ for $a>>a_0$ and $\mu = \frac{a}{a_0}$ for $a<<a_0$ for the recovery of the Newtonian dynamics in the regime where the field is strong enough. For a review \cite{mil13} is useful.\par
In the context of Verlinde's formalism, gravity theories have been connected with models of solid state physics, like the Debye's model at low temperature \cite{refadd1}. In \cite{refadd2} the one dimensional Debye model is shown to give MOND. Some recent attention also includes the derivation of MOND from the holographic entropy area relation \cite{ran1} and the collective motion of holographic screen bits \cite{refadd3}. Here bits are related to the units of information on the holographic screen. In the critical phenomena of cooling it can be shown that if in the equipartition relation, the zero energy bits are removed from the total number, then we can get the notion of MOND. But we have to consider a modified equipartition theorem with the assumption that the division of energy is not homogeneous on all bits below a critical temperature. Then along with the holographic principle and the Unruh temperature we can recover the theory of MOND \cite{s42}. In the language of critical phenomena this is analogous to the first order phase transition. Following the methods of \cite{s42} we consider the fraction of bits with zero energy with 
\begin{equation}
N_0 = N \left(1 - \frac{T}{T_c}\right) ~~.
\end{equation}
So for $T\geq T_c$ there are no bits with zero energy and the zero energy phenomena starts for $T<T_c$. This is a relation for critical phenomena in second order phase transition. The number of bits with different energy at $T<T_c$ is given by
\begin{equation}
N - N_0 = N \left(\frac{T}{T_c}\right) ~~.
\end{equation}
With the equipartition law of energy we get 
\begin{equation}
E = \frac{1}{2}\left(N\frac{T}{T_c}\right)T
\end{equation}
where we have considered $k_B=1$. Now with $E=Mc^2$ we get 
\begin{equation}
\label{e1}
T^2 = \frac{2Mc^2T_c}{N}
\end{equation}
where $M$ is the emergent mass which can be considered to be at the center of the space enclosed by the holographic screen. Now we have the Unruh temperature ($T=\frac{\hbar a}{2\pi c}$) which is associated with the acceleration of the frame and with eqn.(\ref{e1}) we get 
\begin{equation}
\label{e4}
N a^2 = \frac{8\pi^2 c^2}{\hbar^2}Mc^2T_c ~~.
\end{equation}
We have discussed earlier how entropy is related to $N$ as entropy is proportional to the number of bits. Here we use the entropy corrected relation of (\ref{ent}) and modifying $N=4S$ we write
\begin{equation}
\label{eN}
N = \frac{A}{l_p^2} + 4\alpha \sqrt{\frac{A}{4l_p^2}} + 4\beta \ln \frac{A}{4l_p^2} + 4\gamma \left(\frac{A}{4l_p^2}\right)^{-1/2}
+ 4\delta \left(\frac{A}{4l_p^2}\right)^{-1} ~~.
\end{equation}
With $A=4\pi R^2$ and using (\ref{eN}) we re-write eqn.(\ref{e4}) as 
\begin{equation}
a^2 \left(\frac{4\pi R^2}{l_p^2}\right) \left[ 1 + \frac{\alpha ~l_p}{\sqrt{\pi}R} + \frac{\beta ~l_p^2}{\pi R^2} \ln\left\{\frac{\pi R^2}{l_p^2}\right\} + \frac{\gamma ~l_p^3}{\pi^{3/2}R^3} + \frac{\delta ~l_p^4}{\pi^2 R^4} \right] = \frac{8\pi^2 c^2}{\hbar^2} Mc^2 T_c ~~.
\end{equation}
We also mention that we considered only the leading order terms in the entropy area relation of (\ref{ent}). With a little algebra and considering $a_0=\frac{2\pi c}{\hbar}T_c$ we can finally arrive at
\begin{equation}
a\left(\frac{a}{a_0}\right) = \frac{GM}{R^2} \left[ 1 - \frac{\alpha ~l_p}{\sqrt{\pi}R} - \frac{\beta ~l_p^2}{\pi R^2} \ln\left\{\frac{\pi R^2}{l_p^2}\right\} - \frac{\gamma ~l_p^3}{\pi^{3/2}R^3} - \frac{\delta ~l_p^4}{\pi^2 R^4} \right] ~~.
\end{equation}
Here we have only first order terms of $\alpha,~\beta,~\gamma$ and $\delta$. This equation is the entropy corrected equation for the modified Newtonian dynamics.

\section{Entropic Corrections due to GUP and the Newton's Law of Gravitation}

Bekenstein's entropy area relation \cite{ha2} came from the argument that if a particle is within the Compton wavelength from a black hole horizon then it is a part of the black hole. There will be an increase in mass and area of the black hole and the relevant change is identified with one bit of information. With this motivation Verlinde postulated that the entropy associated with the information at the boundary is given by
\begin{equation}
\label{es1}
\Delta S = 2\pi   ~~~~~~~~~~~~~~~~~~\text{when}~~~~~~~~~~~~~~~~~~ \Delta x=\frac{\hbar}{mc} ~~.
\end{equation}
Here we have considered $k_B=1$. If we assume that the change in entropy is linear to $\Delta x$ then we can re-write eqn.(\ref{es1}) as
\begin{equation}
\label{f1}
\Delta S = 2\pi \frac{mc}{\hbar} \Delta x ~~.
\end{equation}
This idea is analogous to osmosis across a semi-permeable membrane. As the membrane carries a temperature $T$ so the particle will experience an effective entropic force
\begin{equation}
\label{f2}
F \Delta x = T \Delta S ~~.
\end{equation}
This force is attractive. A non-zero force leads to a non-zero acceleration and acceleration is related to temperature by Unruh effect. If we now assume that the total energy $E$ of the system is divided evenly over $N$ bits then the temperature is given by the equipartition law of energy
\begin{equation}
T = \frac{2E}{N} ~~.
\end{equation}
With $E=Mc^2$ we get 
\begin{equation}
\label{f3}
T = \frac{2Mc^2}{N} ~~.
\end{equation}
So with (\ref{f1}), (\ref{f2}) and (\ref{f3}) we have 
\begin{equation}
F = \frac{2Mc^2}{N} \frac{2\pi m c}{\hbar} ~~.
\end{equation}
As mentioned earlier we will study the entropy corrected version of this equation. So with the entropy corrections which are incorporated in $N$ (\ref{eN}) we can write the entropic force equation as 
\begin{equation}
F = \frac{GMm}{R^2} \left[ 1 - \frac{\alpha ~l_p}{\sqrt{\pi}R} - \frac{\beta ~l_p^2}{\pi R^2} \ln\left\{\frac{\pi R^2}{l_p^2}\right\} - \frac{\gamma ~l_p^3}{\pi^{3/2}R^3} - \frac{\delta ~l_p^4}{\pi^2 R^4} \right]
\end{equation}
with $A=4\pi R^2$. If $\alpha=\beta=\gamma=\delta=0$ this is the Newton's law of gravitation. The Newtonian potential turns out to be
\begin{equation}
\label{npot}
V(R) \sim \frac{GMm}{R}\left[1 - \frac{\alpha l_p}{2\sqrt{\pi}R} - \frac{2\beta l_p^2}{9\pi R^2} - \frac{\beta l_p^2}{3\pi R^2}
\ln \left(\frac{\pi R^2}{l_p^2}\right) + {\cal O}(l_p^3) \right] ~~.
\end{equation}
This modification of Newton's law is similar to what the predictions came from Randall-Sundrum II model \cite{ali36}. The model has one uncompactified dimension with length scale $l_{\mu}$. But here the sign of the prefactor of the correction is different. If we would have considered the entropy corrected relation of \cite{new} then this sign ambiguity would not have come. In the RS braneworld scenario the Newtonian potential is calculated as \cite{ali37}
\begin{equation}
\label{mGB2}
V(r) \sim \left\{
\begin{array}{ll} 
\frac{1}{R}[1 + \frac{4l_{\mu}}{3\pi R} - \ldots ] ~~~~~~~~~~~~ \text{for} ~~~~~~~~ l_{\mu}>> R   \\\\
\frac{1}{R}[1 + \frac{2l_{\mu}^2}{3 R^2} - \ldots ] ~~~~~~~~~~~~~ \text{for} ~~~~~~~~ l_{\mu}<< R
\end{array} \right. 
\end{equation}
where $l_{\mu}$ is the characteristic length scale of the theory. The significant prediction of \cite{ali37} is that gravity is five dimensional at short distances. This comparison of the modified Newton's law of gravitation with respect to the RS II model was first pointed out in \cite{ali}. The modified entropy-area relation that we have used is a consequence of the GUP and the question can be raised that whether the GUP modifications predict same results as that of RS II model for short distance physics. Although Newton's $\frac{1}{R^2}$ force law is the only law of gravitation upto 0.13 mm \cite{ali56} but still it is unknown whether the law is valid at much lower scales. Here it is possible to put an upper bound on $\alpha$ from the RS II characteristic length scale $l_{\mu}$.
If the tension $\frac{1}{l_{\mu}}$ of the brane is small enough compared to the Planck mass then the correction to the Newtonian potential would help us to distinguish RS II model with other extradimensional models. $l_{\mu}$ is constrained by present short
distance tests of gravity which says $l_{\mu} < 11 \times 10^{-06}$ m \cite{palma4}. If we use this bound on $l_{\mu}$ and compare with the first order correction of (\ref{npot}) we get an upper bound on the deformation parameter $\alpha$ which is $<10^{29}$.
This bound is not sensitive for phenomenological purposes as this intermediate length scale should be $\leq 10^{17}$ otherwise it
would have been observed as the electroweak length scale is $\sim 10^{17}l_p$ \cite{new31}. Also the current experimental bounds on $\alpha$ are $\leq 10^{17}, 10^{10}, 10^{11}$ from position measurement, Hydrogen Lamb shift and electron tunneling respectively \cite{add45,p20}. On the other hand in 5-D Heterotic M-theory if the 5-dimensional fundamental mass is of the order of grand unification scale ($10^{16}$GeV) then the corrections to the Newtonian potential would be relevant at $l_{\mu} \sim 10^{-26}$m \cite{palma}. Now correction of this order if compared to the first order correction of eqn. (\ref{npot}) would give an upper bound
for the parameter $\alpha$ which is $<10^{9}$.\par
It is also interesting that if we consider the scattering of two heavy masses $m_1$ and $m_2$ in a gravitational potential, the non
relativistic potential get some corrections. We can write
\begin{equation}
\label{dono}
V(r) \sim \frac{Gm_1 m_2}{r} \left[1 + \frac{3G(m_1 + m_2)}{rc^2} + \frac{41 l_p^2}{10 \pi r^2}\right] ~~.
\end{equation}
This is the Donoghue potential \cite{dono1,dono2}. The first correction term is the classical classical post Newtonian correction and the last correction is purely quantum. For the derivation one has to treat quantum general relativity as an effective field theory. The last correction term of the Donoghue potential do not fit well for the phenomenological purposes but it definitely shows a well
behaved classical limit. The first correction term of the potential is not considered as a quantum correction as for a small test particle $m_2$ this is similar to the time component $g_{00}$ of the Schwarzschild metric which is the source of the static gravitational potential \cite{dono2}. In the process to get eqn. (\ref{npot}) Newton's law arose naturally with the consideration that space is emergent through a holographic scenario \cite{s14}. But eqn. (\ref{dono}) is an artifact of treating quantum general relativity as an effective field theory. Here also we see that the predictions of an emergent theory of gravity with the effects of the generalized uncertainty principle are similar to that of quantum general relativity. Although this is not surprising that any theory of quantum gravity comes with an intrinsic length scale and the low energy effective theory is plagued with corrections associated with the length scale.

\section{Entropic Corrections due to GUP and the Einstein's Equation}

In the earlier section we have considered non homogeneous cooling of bits which restricted the distribution of energy equally on all bits of the holographic screen below a critical temperature. In turn we get a modified equipartition law of energy and we derive MOND in presence of the effects of generalized uncertainty principle. In this section we further investigate the effects of the results derived earlier on the Einstein's field equations. With the assumption that the holographic principle holds and considering the fact that each single bit of information occupies a unit cell one can write
\begin{equation}
N = \frac{A}{l_p^2} ~~~~~~~~~~~~~\text{where}~~~~~~~~~~~~~~~l_p^2 = \frac{G\hbar}{c^3} ~~.
\end{equation}
This is four times the Bekenstein-Hawking entropy which says $S_{BH} = \frac{A}{4l_p^2}$. So that we can write
\begin{equation}
N = 4S
\end{equation}
Considering the entropic corrections due to GUP (\ref{eN}) we can write
\begin{equation}
dN = \frac{1}{l_p^2} \left[1 + \frac{\alpha~l_p}{\sqrt{A}} + \frac{4\beta ~l_p^2}{A} - \frac{4\gamma ~l_p^3}{A^{3/2}} - \frac{16\delta ~l_p^4}{A^2} \right]dA ~~.
\end{equation}
This is the bit density on the screen. If the energy associated with mass $M$ is divided over all bits and each bit carries mass $\frac{1}{2}T$ due to the equipartition law we have 
\begin{equation}
\label{e20}
M = \frac{1}{2} \int_{{\cal S}} T ~dN ~~.
\end{equation}
The local temperature $T$ on the screen is given by
\begin{equation}
T = \frac{\hbar}{2\pi} e^{\phi}n^b \nabla_b \phi
\end{equation}
where $e^{\phi}$ is the redshift factor as $T$ is measured from infinity. So eqn.(\ref{e20}) is written as
\begin{equation}
\label{e25}
M = \frac{1}{4\pi G} \int_{{\cal S}} e^{\phi}~\nabla \phi \cdot \left[1 + \frac{\alpha~l_p}{\sqrt{A}} + \frac{4\beta ~l_p^2}{A} - \frac{4\gamma ~l_p^3}{A^{3/2}} - \frac{16\delta ~l_p^4}{A^2} \right]dA
\end{equation}
where we have considered $c=1$. This equation is the modified Gauss Law in general relativity and the right hand side represents the modified Komar mass. The first integral of eqn.(\ref{e25}) is the Komar mass $M_K$ and 
\begin{equation}
M_K = \frac{1}{4\pi G} \int_{{\cal S}} e^{\phi}~\nabla \phi \cdot dA ~~.
\end{equation}
Now this relation can be written in terms of the Ricci tensor $R_{ab}$ and the Killing vector $\xi^a$ \cite{s44,s14} where one uses the Stokes theorem and the Killing equation for $\xi^a$: $\nabla^a \nabla_a \xi^b = -{R^b}_a\xi^a$. Finally one can get
\begin{equation}
M_K = \frac{1}{4\pi G} \int_{\Sigma} R_{ab}~n^a~\xi^b ~dV ~~.
\end{equation}
So we rewrite eqn.(\ref{e25}) as
\begin{equation}
\label{e26}
M = \frac{1}{4\pi G}\int_{\Sigma} R_{ab}~n^a~\xi^b ~dV + \frac{l_p}{4\pi G} \int_{{\cal S}} e^{\phi}~\nabla \phi \cdot \left[ \frac{\alpha~}{\sqrt{A}} + \frac{4\beta ~l_p}{A} - \frac{4\gamma ~l_p^2}{A^{3/2}} - \frac{16\delta ~l_p^3}{A^2} \right]dA
\end{equation}
where $\Sigma$ is the three dimensional volume bounded by ${\cal S}$ which is the holographic screen and $n^a$ is the normal. Also $M$ can be written as a volume integral of the stress energy tensor $T_{ab}$ where 
\begin{equation}
\label{e27}
M = 2 \int_{\Sigma} \left(T_{ab} - \frac{1}{2} T g_{ab}\right) n^a ~\xi^b ~dV
\end{equation}
So with eqn.(\ref{e26}) and eqn.(\ref{e27}) we can write the entropy corrected Einstein's equation as 
\begin{align}
\int_{\Sigma}\Big[R_{ab}  & - 8\pi G \Big(T_{ab} - \frac{1}{2}Tg_{ab}\Big)\Big]n^a~\xi^b~dV  \nonumber \\
& = -l_p \int_{{\cal S}} e^{\phi}~\nabla \phi \cdot \left(\frac{\alpha~}{\sqrt{A}} + \frac{4\beta ~l_p}{A} - \frac{4\gamma ~l_p^2}{A^{3/2}} - \frac{16\delta ~l_p^3}{A^2} \right)dA ~~.
\end{align}
If $\alpha=\beta=\gamma=\delta=0$ we get the usual Einstein's equation. Here we have surface corrections which came as a consequence of the correction to the density of bits on the holographic screen. In a spherically symmetric static space time with a little algebra finally we can get the entropy corrected Einstein's equation as 
\begin{equation}
R_{ab} = 8\pi G \Big(T_{ab} - \frac{1}{2} T g_{ab}\Big) (1 + \alpha')
\end{equation}
where 
\begin{equation}
\alpha' = \frac{l_p}{2\pi} \left(\frac{\alpha~}{\sqrt{A}} + \frac{4\beta ~l_p}{A} - \frac{4\gamma ~l_p^2}{A^{3/2}} - \frac{16\delta ~l_p^3}{A^2} \right) ~~.
\end{equation}
For large horizon area the equation reduces to usual Einstein's equation.

\section{Discussion}

Gravity may be identified to be associated with entropic force and a thermodynamical system may well describe a gravitational system. This idea came from the thermodynamical interpretation of gravitational field equations. A holographic screen is assumed to contain the information of the volume enclosed by it and the information is divided in bits. So according to Verlinde it is natural to assume that the number of bits is proportional to the area of the holographic screen. On the other hand, all approaches to quantum gravity support the idea of existence of a minimal observable length of the order (or some order) of Planck length. Also it is conjectured that the standard commutation relations at short distances would be modified. In \cite{new31} a form of the generalized uncertainty principle was proposed which is consistent with Doubly Special Relativity (DSR), string theory and black holes physics with the prediction of a maximum observable momentum and a minimal measurable length. As an immediate effect of the quantum gravity corrections incorporated through the GUP the entropy area relation of a black hole gets modified. So it leads to a
modification in the number of bits of information on the holographic screen as discussed in Verlinde's approach. Verlinde's approach is found to be consistent if one obtain modifications to the Newton's law from the log corrected entropy area relation as the modifications has the same form as that of lowest order quantum corrections of perturbative quantum gravity \cite{ran1}.\par
In this paper we have generalized the entropic force law as introduced by Verlinde via a phenomenological interpretation of the generalized uncertainty principle. Considering the effects of generalized uncertainty principle in the black hole entropy area relation here we have derived the modified equations for Newton's law of gravitation, modified Newtonian dynamics and Einstein's general relativity. The leading order correction of the modified potential in the Newton's Law of gravitation surprisingly agrees with the short distance Newtonian potential as predicted by Randall-Sundrum II model. As the RS II model has an uncompactified extra dimension so it would be interesting to investigate whether the GUP effects can predict the same as that of the extra dimensional theories. It is also interesting to note that the corrections to the Newtonian potential which we get are similar to the Donoghue potential which is a consequence of treating quantum general relativity as an effective field theory. We found that the corrections due to an emergent theory of gravity with the effects of minimal length are similar to that of quantum general relativity. This is quite evident as any theory of quantum gravity is accompanied by an intrinsic length scale which manifests itself as the coefficient of leading order corrections in low energy phenomena. Here with Verlinde's approach we observe that the GUP motivated entropy area relation modifies Newton's law with modifications that are similar to different order of quantum effects as evidenced in perturbative quantum gravity. Later we derived the modified Einstein's field equation in the same framework which for large horizon areas reduces to the usual Einstein's field equation.

%End
\subsubsection*{Acknowledgments}
The author would like to acknowledge the support of APS-IUSSTF Visitation Program as a part of the present work was completed during the visit. The author would also like to thank an anonymous referee for enlightening comments and helpful suggestions.

%\section*{References}

\end{document}